\documentstyle[12pt,twoside,fleqn,espcrc1,epsfig]{article}
\newcommand{\AmS}{{\protect\the\textfont2
  A\kern-.1667em\lower.5ex\hbox{M}\kern-.125emS}}

\title{Associated Strangeness Production in the 
       Threshold Region}

\author{W. Oelert\thanks{Talk presented at the conference HYP'97, BNL, October 1997. \protect \\
The author would like to thank the members of
both collaborations PS185 at LEAR/CERN and COSY-11 at COSY-J\"ulich for
many exciting hours of data taking, data evaluation, discussions and
interpretations. Thanks are due to John Millener for his careful reading of the manuscript.}
\address{Institut f\"ur Kernphysik, 
         Forschungszentrum J\"ulich, \\
         P.O. Box 1913,  D - 52425 J\"ulich, Germany}%
           }
        
\begin{document}
\maketitle 

\begin{abstract}Associated strangeness production close to threshold is a 
particularly clean example of how energy is converted into hadronic matter. 
Final results from the completed experiments of the PS185 \mbox{collaboration at 
LEAR/} CERN, using the antiproton-proton interaction to produce antihyperon-hyperon
pairs, demonstrate that such studies form a powerful tool. 
First results from the ongoing investigations of the COSY-11 
collaboration at COSY-J\"ulich, measuring the strangeness dissociation into both
hyperon-kaon and kaon-kaon pairs from proton-proton scattering, are
presented and plans for further investigations are discussed.
Finally, an interesting possibility for future studies of the strangeness
production mechanism
in the antiproton-proton interactions with antiproton beams of momenta $\ge
5~GeV/c $ is suggested.   
\end{abstract}

\section{INTRODUCTION}
The focus of the experiments discussed in this contribution is to explore the
physics of strange-antistrange quark production and the role of these quarks
in the configuration of the emerging hyperons and/or mesons. In view of the
attempts to understand the nature of the quark-flavour composition, its
structure,
and its creation process, the associated strangeness production is of fundamental
interest.
It is essential that different complementary tools are used for the
understanding of the complex hadronic systems. The first results on electromagnetic
probes have been reported during the present conference. However, we concentrate on 
the hadronic entrance channel interaction.\\
\\
In a systematic study, the PS185 collaboration at the Low-Energy Antiproton Ring
at CERN (LEAR/CERN) investigated antihyperon-hyperon $(\bar Y Y)$
production and the decay via the reaction $\bar p p \rightarrow \bar Y Y
\rightarrow \bar p \pi^+~p \pi^- $ in the threshold region \cite{JOH97}. Cross
sections, 
polarization, spin correlations and singlet fractions have been extracted.\\
At the COler SYnchrotron COSY-J\"ulich
the COSY-11 collaboration investigated the associated strangeness production via
two channels. In the  $p p \rightarrow p K \Lambda $ reaction
\cite{BAL96,BAL97} the strangeness is dissociated into a hyperon and a meson and in
the $p p \rightarrow p p K^+ K^- $ reaction \cite{WOL97,LIS98} the 
antistrange-strange quark pair is divided between two mesons. Total cross sections 
have been determined and lead to the first physics interpretations.\\
\\
Theoretical models in nuclear physics, in particular heavy-ion physics, depend critically on
input from elementary particle interactions. Meson-nucleon and meson-meson
interactions dominate the nucleus-nucleus scattering mechanisms. Therefore, knowledge of
the kaon production cross section in the elementary NN interaction is important
for studies of the production of hypernuclei in nucleon scattering on nuclei,
and for investigations of the strangeness production mechanism in heavy-ion
collisions, which may provide information about hot, dense nuclear matter and
eventually the existence of a quark-gluon plasma. More details and
applications are given in the contribution by G. Brown in these proceedings.\\
\\
Both experimental investigations - PS185 and COSY-11 - have been carried out in the 
production threshold region offering the advantage that \\
1) experimentally the reaction products are confined in a forward cone, whereas \\
$~~~~~$background reactions populate the phase-space more homogeneously and \\
ii) the momentum transfers are relatively large with only the lowest partial waves \\
$~~~~~$contributing, which should simplify the theoretical framework.\\
\\
Finally, a natural continuation of the associated strangeness production would
be a study and a comparison of the two systems $\bar \Omega \Omega$ and 
$\phi \phi \phi$ \cite{DOV91,OEL91,OEL93}, which both contain three $s$ quarks and 
three $\bar s$ quarks in different configurations at rather similar masses as 
3.344 GeV/$c^2$ and 3.058 GeV/$c^2$, respectively.

\subsection{The two Accelerators LEAR and COSY}

The LEAR cooler and accelerator \cite{CES91} might well be regarded as the
father of 
a whole generation of further similar machines as the cooler at Indiana
\cite{POL91}, the 
CELSIUS ring in Uppsala \cite{REI97}, and the COSY facility at J\"ulich
\cite{MAI97}. 
Whereas LEAR, with an antiproton beam
of momenta $\le 2~GeV/c$, was shut down at the end of 1996 after 15 years of
very successful operation, COSY is just at the beginning of an extensive program
using proton beams with momenta $\le 3.4~GeV/c$. Polarized proton beams are
presently
under development. Both machines have electron as well as stochastic cooling
facilities, resulting in a high phase-space densitiy with momentum resolution of
\mbox{$\Delta p/p \approx 10^{-4}$}. At both accelerators, internal and external
target stations were/are used.\\
For the $\bar p p$ interaction at LEAR a total energy of 
$\sqrt s ~=~ 2.43~GeV$ was available at the maximum beam momentum and, due to
the baryon number B~=~0, this energy could entirely be converted into newly
created hadronic matter. At COSY the $p p$ interaction results in a maximum
total
energy of $\sqrt s ~=~ 2.86~GeV$ but, due to conservation of the baryon number B~=~2,
the energy which may be converted into hadronic matter is about
1~GeV.\\
In order to demonstrate the complementarity of the two accelerators and their 
experiments, a brief summary of results from one experiment at LEAR will be given next, 
before turning to the first results from COSY.

\section{$\bar Y - Y$ PRODUCTION in PS185}

\begin{figure}[htb]
\begin{minipage}[t]{70mm}
\epsfig{file=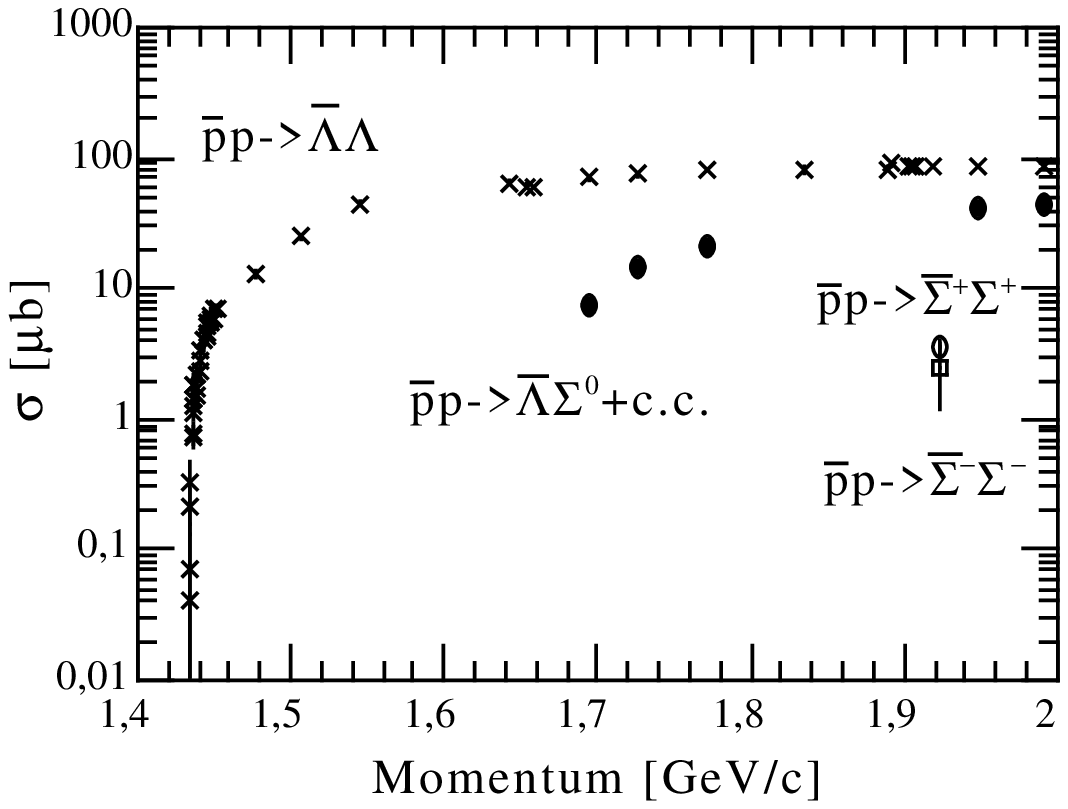,width=75mm}
\caption{Total cross sections for:  \protect \\
$~~~~~~~~~~~~~\bar p p \rightarrow \bar \Lambda \Lambda ~~~~~~~~~$(crosses) \protect \\
$~~~~~~~~~~~~~\bar p p \rightarrow \bar \Lambda \Sigma^0 + cc~~$(filled circles) \protect \\
$~~~~~~~~~~~~~\bar p p \rightarrow \bar \Sigma^+ \Sigma^+~~~~~~$(open circles) \protect \\
$~~~~~~~~~~~~~\bar p p \rightarrow \bar \Sigma^- \Sigma^-~~~~~~$(open square)}
\label{excitation}
\end{minipage}
\hfill
\begin{minipage}[t]{70mm}
\epsfig{file=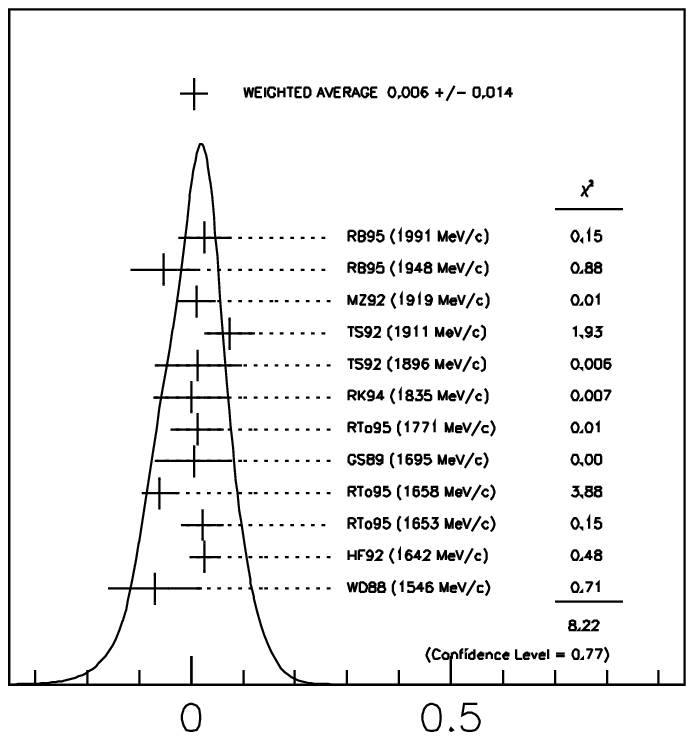,width=70mm}
\caption{Data for the parameter A \protect \\
$~~~~~~~~~~~~~~$as collected by the PS185 \protect \\
$~~~~~~~~~~~~~~$collaboration.
}
\label{asym}
\end{minipage}
\end{figure}

The PS185 collaboration \cite{JOH97} has collected high-precision data on the
$\bar p p \rightarrow \bar Y Y$ reactions at LEAR, with the following conclusions.\\
The excitation functions of the $\bar \Lambda \Lambda$ and 
$\bar \Lambda \Sigma^0 + cc $ productions follow a similar pattern, as can be
seen in Fig.~\ref{excitation}. The measured cross sections for the charged 
$\Sigma$ hyperons do not agree with predictions based on quark-line diagrams but
rather seem to favour a meson-exchange picture \cite{HAI92}.\\
An early onset of p-wave contributions to the cross section is found near
threshold. An universal behaviour of the $\bar Y Y$ differential cross sections,
showing a strong forward rise followed by a flat distribution, is observed
except for the $\bar \Sigma^- \Sigma^- $ channel. The shapes of the 
$\bar \Lambda \Lambda$ differential cross sections and the spin correlations 
appear to be essentially energy independent which is not the case for the 
$\Lambda (\bar \Lambda$) polarisation. The $\bar \Lambda \Lambda$ pairs are
produced in parallel-spin configurations with a weighted mean singlet fraction
of 
\verb=<=$S_F\verb=>==0.007\pm0.009$, whereas the spins of the 
$\bar \Lambda \Sigma^0 + cc$ channel show a more statistical distribution.
Finally, the normalized asymmetry value 
$A~=(\alpha + \bar \alpha)/(\alpha - \bar \alpha)$ (where $\alpha (\bar \alpha)$
is the asymmetry parameter for  $\Lambda (\bar \Lambda)$)
has a value of $0.006~\pm~0.014$ (see Fig.~\ref{asym}), which is, at least on
this level, consistent with zero, as expected if CP is conserved.
 
\section{Associated Strangeness Production in the pp Scattering}

\begin{figure}[htb]
\centerline{\epsfig{file=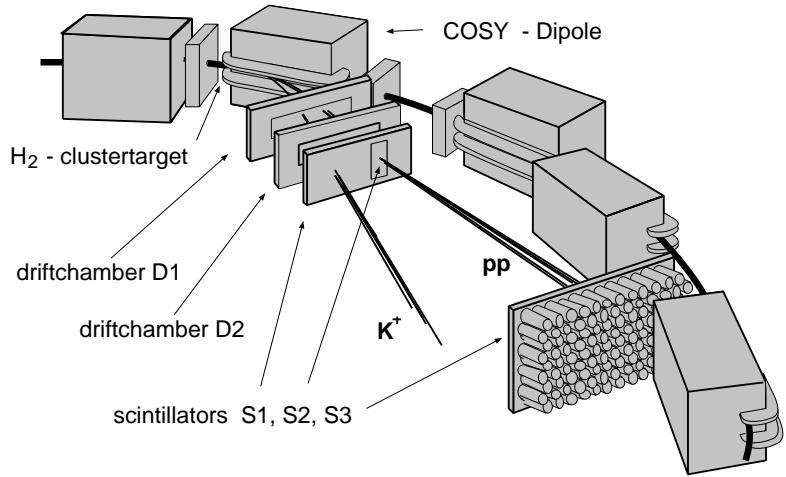,width=0.7\textwidth}}
\caption{The COSY-11 installation in a bending section of the COSY ring. The \protect \\
$~~~~~~~~~~~~~~~$ejectile momenta are reconstructed using the track information
from: \protect \\ 
$~~~~~~~~~~~~~~~$the drift chambers D1, D2 and time of flight from the
scintillator \protect \\
$~~~~~~~~~~~~~~~$hodoscopes S1, S2 and S3. The plotted tracks result from MC
calcu- \protect \\ 
$~~~~~~~~~~~~~~~$lations of some $pp \rightarrow pp K^+ K^-$ events at an excess
energy of 2 MeV.}
\label{c11detsw}
\end{figure}

The first results of elementary reactions have been obtained with COSY: \\
i) at the installation COSY-11, measurements for the processes:
\mbox{$pp \rightarrow pp \eta $},
\mbox{$pp \rightarrow pp \eta '$},
\mbox{$~~~pp \rightarrow pp K^+ K^-$},
\mbox{$pp \rightarrow p K^+ \Lambda$}, and
\mbox{$pp \rightarrow p K^+ \Sigma^0~~$}  \\
ii) at the facility TOF, measurements for the\mbox{ $pp \rightarrow p K^+ \Lambda$} reaction 
at higher excess \\
$~~~~$energies.\\
\\
The COSY-11 installation \cite{BRA96} is an internal setup in a bending section
of the COSY ring. 
Figure \ref{c11detsw} shows a sketch of the experiment. 
In front of a machine dipole, a hydrogen cluster target \cite{DOM97} is
installed. 
Positively charged reaction products - which have a reduced momentum 
compared to the primary proton beam - are bent to the inner region 
of the ring and are detected with a system of drift chambers and scintillators.
Negatively charged ejectiles are detected by a silicon pad detector 
arrangement which follows a long scintillation detector in the dipole gap; 
these counters are not seen in the figure.

\subsection{Strangeness Production into $p K^+ \Lambda$}
  
\begin{figure}[htb]
\centerline{\epsfig{file=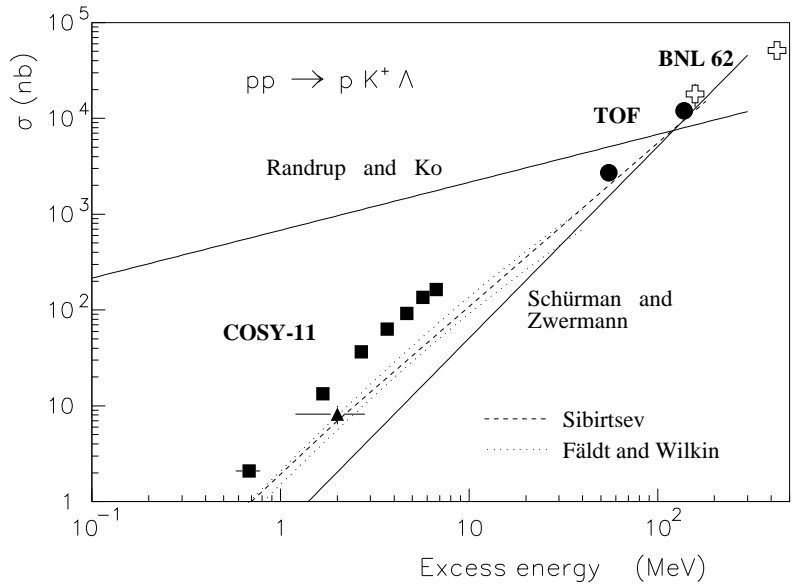,width=0.7\textwidth}}
\caption{Existing data set in the threshold 
region for the $pK^+ \Lambda$ production.}
\label{pkldt}
\end{figure}
In Fig. \ref{pkldt}, the world set of data for the 
$pp \rightarrow p K^+ \Lambda$ reaction up to an excess energy of 200 MeV
is shown. Before the startup of COSY, the lowest-energy data point for the total 
production cross section was at $\approx$ 160 MeV excess energy \cite{FIC62}.
(Inclusive $K^+$ production for a limited angular range was performed at SATURNE for 
lower excess energies \cite{FRA89}.) In the threshold region, several data points have been 
added by experiments at COSY, where the result at the highest momentum is consistent
with the lowest bubble chamber data \cite{FIC62}. The lines represent different model 
calculations. The parametrisations of Randrup and Ko \cite{RAN80} and of Sch\"urmann 
and Zwermann \cite{ZWE88} (solid lines), and new calculations of F\"aldt and Wilkin \cite{FAL97}
(range between dotted lines) and Sibirtsev \cite{SIB95} (dashed line), are shown.\\ 
A crucial requirement for measuring cross sections in the threshold region is an exact 
knowledge of the beam momentum, which is the dominant error for the excitation curve.
The COSY-11 data were used by themselves for the calibration of the beam momentum. The 
absolute momentum setting for a fixed COSY optics is limited to the order of $10^{-3}$ 
due to an insufficient knowledge of the orbit length of the COSY beam.
The relative error for a momentum change at the same optics, however, is negligible. 
Therefore a data set taken in one run with a fixed optics can be used for the 
calibration. The data are fitted by a phase-space distribution including 
$p\Lambda$ final-state interaction and the Coulomb correction for the 
$pK^+$ system, from which the correct excess energy could be extracted. The deviation
from the nominal excess energy given by COSY was as little as 0.2 MeV
\cite{BAL97}.

\subsection{Strangeness Production into $pp~ K^+ K^-$}
  
The production of a $K^+ K^-$ meson pair is very interesting 
in view of the existence of $K\bar{K}$ molecules and 
especially for the structure of the objects $f_0$(980) and $a_0$(980). 
The structure of these two
resonances has provoked a long standing discussion in the literature describing them as
normal $q \bar{q}$ states, exotic hybrids, hadronic $K^+ K^-$ molecules and even
glueball candidates. Again, they have been at the center of attention during the
recent
HADRON'97 \cite{HAD97} conference where the controversial discussion could best
be summarized by three presentations given\\
\\
i)\hspace{0.5cm}\begin{minipage}[t]{14cm}
by Lafferty \cite{HAD97} from the OPAL collaboration, which observed the decay
$Z^0 \rightarrow f_0(980),~f_2(1270)$, and $\phi(1020)$, and
conclude that all features of the $f_0(980)$ are consistent with a pure
$q\bar{q}$ scalar-meson description,
\end{minipage}\\[0.5em]
ii)\hspace{0.4cm}\begin{minipage}[t]{14cm}
by Lebrun \cite{HAD97} from the FERMILAB E-687 collaboration, which found a strong
$f_0(980)$ resonance in the invariant mass of two oppositely-charged pions in
the $D_s \rightarrow \pi^+ \pi^- \pi^+$ decay, but observed no such resonance
in the $D^+ \rightarrow \pi^+ \pi^- \pi^+$ channel, an indication of a significant
strangeness component in the $f_0(980)$ resonance. However, for final
conclusions additional data would be required,
\end{minipage}\\[0.5em]
iii)\hspace{0.3cm}\begin{minipage}[t]{14cm}
by Kirk \cite{HAD97} from the OMEGA collaboration, which measured central proton-
proton collisions and found that "the $f_0(980)$ acts in an opposit
fashion from all known $q \bar{q}$ states". This observation is based on
a small momentum transfer to the produced resonance state, as typically
expected for gluon-rich mechanisms.
\end{minipage}\\[0.5em]
For a general review of the present stage of speculations about the scalar
mesons and glueballs from the theoretical and experimental point of view see 
ref. \cite{CLO97,KLE97}.
\begin{figure}[htb]
\begin{minipage}[t]{70mm}
\epsfig{file=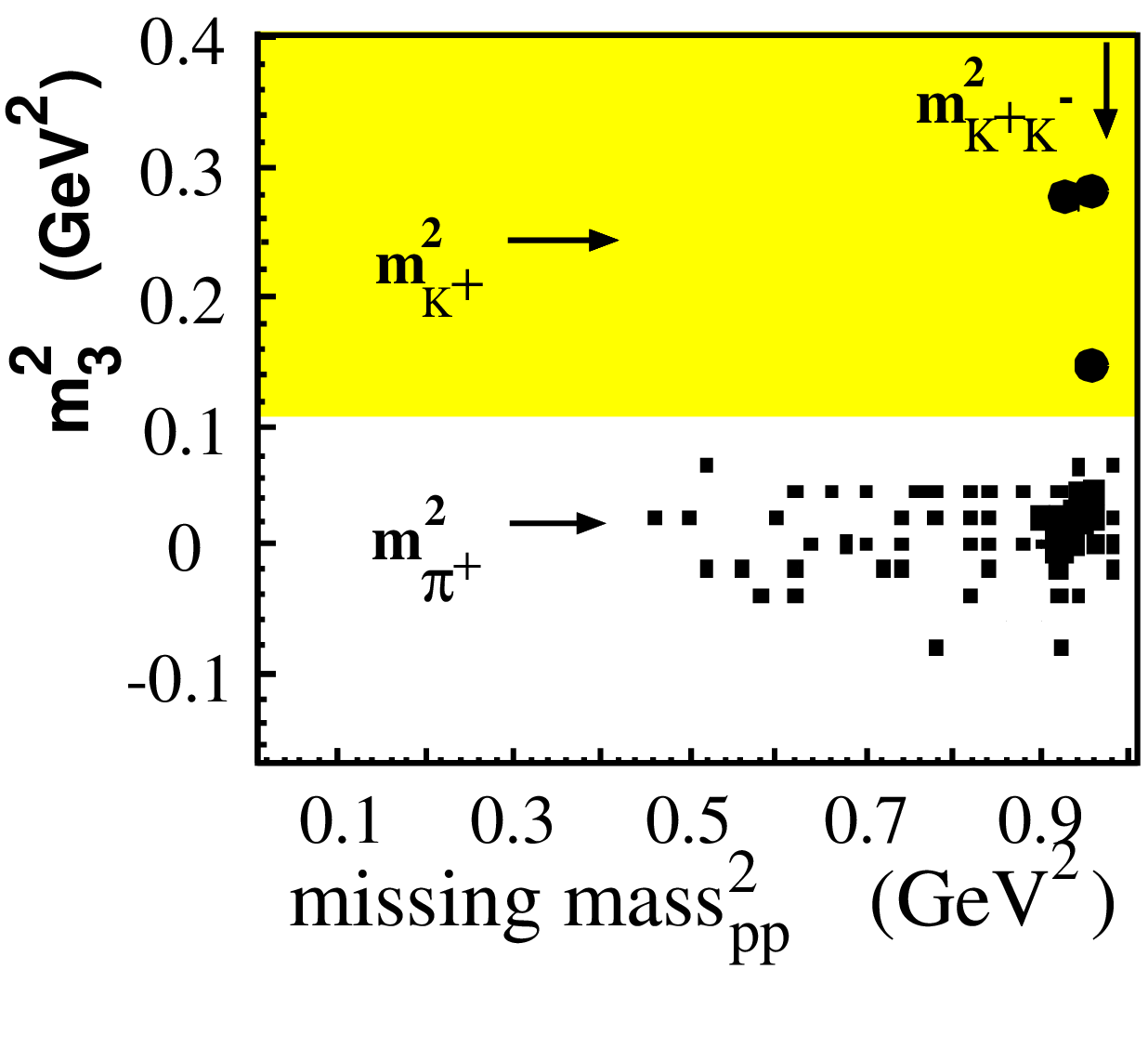,width=70mm}
\caption{Squared invariant mass of \protect \\
$~~~~~~~~~~~~~~$the third charged ejectile  \protect \\
$~~~~~~~~~~~~~~$versus the squared missing   \protect \\
$~~~~~~~~~~~~~~$mass in the pp system for   \protect \\
$~~~~~~~~~~~~~~$data at 3.321 GeV/c  \protect \\
$~~~~~~~~~~~~~~$beam momentum.}
\label{kk321pp}
\end{minipage}
\hfill
\begin{minipage}[t]{70mm}
\epsfig{file=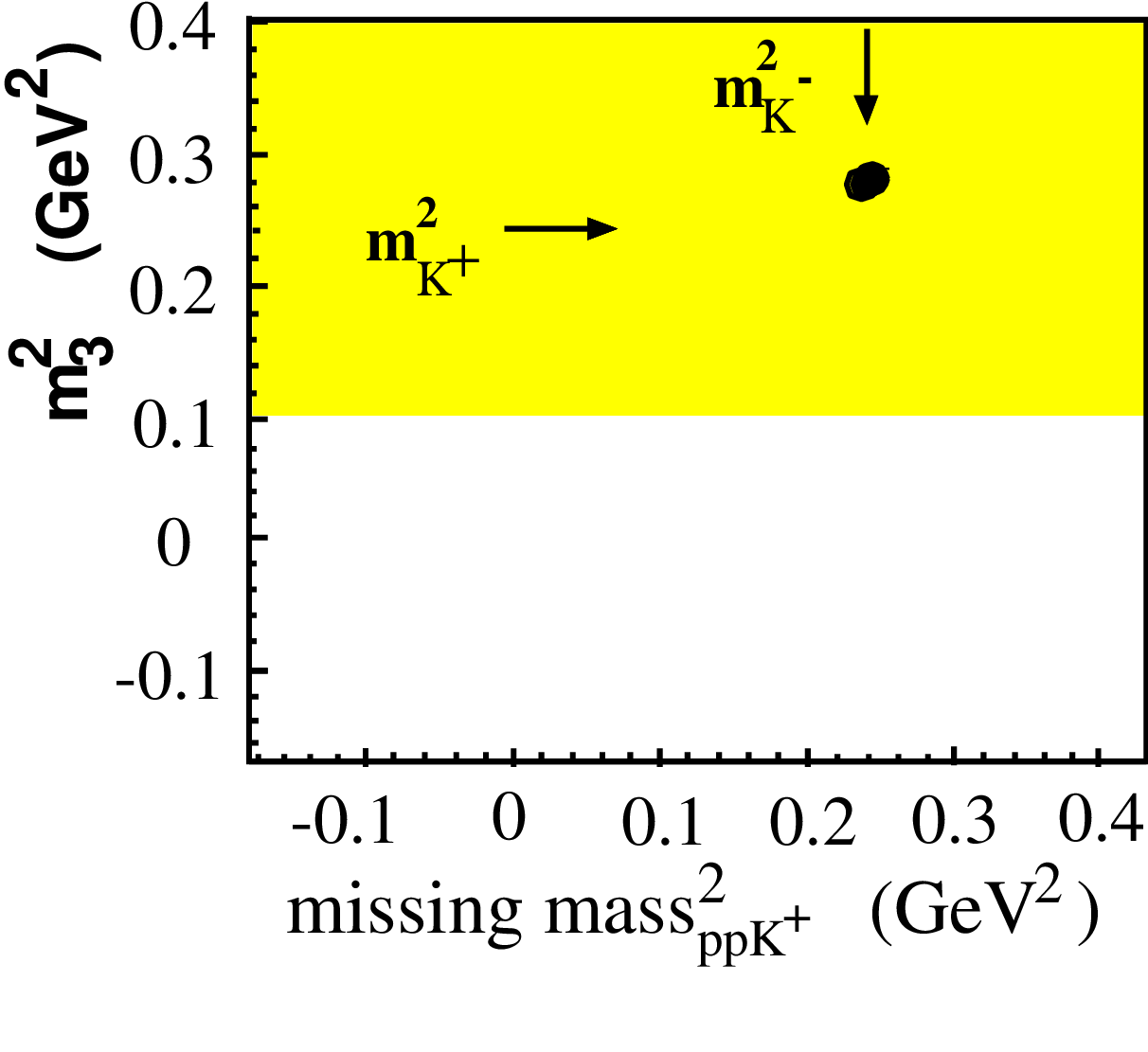,width=70mm}
\caption{Squared invariant mass of   \protect \\ 
$~~~~~~~~~~~~~~~$the third particle vs.   \protect \\ 
$~~~~~~~~~~~~~~~$the missing mass in the   \protect \\
$~~~~~~~~~~~~~~~ppK^+$ system. Two events  \protect \\
$~~~~~~~~~~~~~~~$remain consistent with a   \protect \\
$~~~~~~~~~~~~~~~$$pp \rightarrow ppK^+ K^-$ hypothesis.}
\label{kk321kk}
\end{minipage}
\end{figure}
\\
The first results of the COSY-11 collaboration on $K^+ K^-$ meson pair
production in the 
$pp~\rightarrow~pp~K^+ K^-$ reaction are given in Figs. \ref{kk321pp} and
\ref{kk321kk} \cite{WOL97}, 
which show missing masses vs. the invariant mass of the third positively-charged
particle. 
The gray region indicates the invariant mass range which is accepted for the
kaons in the further 
analysis. The data result from a run at 3.321 GeV/c, which corresponds to
an excess energy of 6.1 MeV. Two events could be selected with a clear signature
for 
the $K^+ K^-$ production. The resulting cross section is $\approx 500 pb$.
In the measurement below threshold,
made with comparable luminosity, no events were observed.\\ 
At COSY-11, further data on $K^+ K^-$ production have been taken at different
excess energies, 
with an expected yield of some ten events for each momentum setting. The data
are under evaluation 
and will give an excitation curve in the threshold region which might serve for
a first 
comparison to theoretical expectations. Preliminary results \cite{LIS98} seem to
indicate that
the relative cross section increase with increasing excess energy (6 $\rightarrow$
27 MeV) is much
smaller than would be expected by a four-body phase-space distribution for the
reaction
$pp~\rightarrow~pp~K^+ K^-$. On an absolute scale, no theoretical cross
section predictions to which the present data could be compared are available.

\section{Triple Associated Strangeness Production}

A systematic study of $\bar Y Y$ pairs with increasing strangeness content is
certainly interesting due to the change of quark dynamics in the different flavour
composition of hadronic matter. As more and more up and down quarks are 
replaced by strange quarks to produce the higher-mass baryons, the importance of
boson-exchange contributions may diminish. At the same time, the gluonic degrees of freedom
may become of increasing significance. The possibility of observing three 
$\bar s$ and three $s$ quarks in different hadronic
environments should be stressed, i.e. the production into the final channels of an
$\bar \Omega \Omega$ baryon pair compared to three $\Phi$ mesons ($\Phi \Phi \Phi$). 
Certainly, a study of such production mechanisms will contribute to the knowledge of 
dynamics in the quark-gluon sector. Angular distributions of the
production of $\bar \Omega \Omega$ baryon pairs could provide further insight.
Whereas $\Phi \Phi$ meson production will always result in
symmetric angular distributions \cite{JETSET} since they are indistinguishable,
the $\bar \Omega$ (or $\Omega$) would be symmetric around $90^{\circ}$ only if the
intermediate state is a compound like gluonic system. If, however, a meson-exchange
process is dominant, such symmetry would most likely be broken.\\
A distinct feature of the $\bar \Omega \Omega$ production is the creation of two
\begin{figure}[htb]
\centerline{\epsfig{file=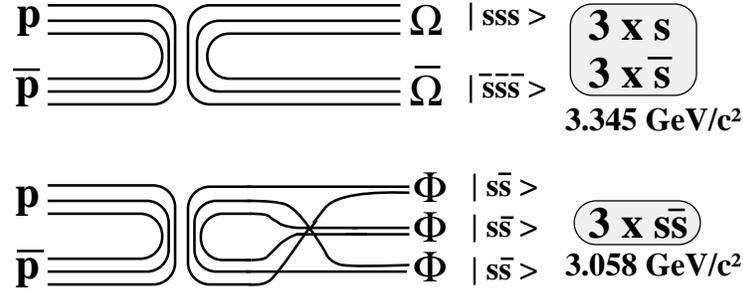,width=0.68\textwidth}}
\caption{Intuitive quark-flow diagram for reaction products following the \protect \\
$~~~~~~~~~~~~~~$$\bar p p$ interaction into the channels:
$\bar \Omega \Omega$, and 
$\Phi \Phi \Phi$. }
\label{omegaphi} 
\end{figure}
spin 3/2 objects. Results from the $\bar \Lambda \Lambda$ PS185 experiment provide
proof of a clear dominance of the triplet $\bar s s$ quark production in the threshold
region. Since in $\bar \Omega$ ($\Omega$) the three $s$ quarks 
(three $\bar s$ quarks) are orientated parallel, the three $\bar s s$ pairs
produced out of the gluonic intermediate state must have spin-parity quantum
numbers $3^-$, whereas in the case of triple  $\Phi \Phi \Phi$ production
such a correlation is not required. Figure \ref{omegaphi} shows a symbolic
representation of a quark-line diagram. It would certainly be very interesting
to observe which type of configuration is preferred by nature. \\
\\
To end by quoting Dover \cite{DOV91} from his conclusions at the Workshop on
Physics at SuperLEAR in October 1991: "Neither of these pictures - meson exchange or quark
mechanisms - offers reliable predictive power, although estimates for
$\bar p p \rightarrow \bar \Omega \Omega$ were presented". Based on the similar
mass values of the two systems, he speculated further that near the 
$\bar \Omega \Omega$ threshold, bound states which decay via
$\bar \Omega \Omega \rightarrow 3 \Phi $ in a natural way may be found.\\
\\
In this contribution, the mutual complementarity of different experiments has been
demonstrated. Using different tools at different accelerators, a unique
understanding of complicated hadronic matter should arise.


\begin{thebibliography}{9}
\bibitem{JOH97} T. Johansson, Nucl. Phys. B \textbf{56A} (1997) 46, and
references therein.
\bibitem{BAL96} J. Balewski et al., Phys. Lett. \textbf{B 388} (1996) 859. 
\bibitem{BAL97} J. Balewski et al., Phys. Lett. \textbf{B}., in press.
\bibitem{WOL97} M. Wolke, PhD thesis, University Bonn, 1997.
\bibitem{LIS98} T. Lister, PhD thesis, University M\"unster, 1998.
\bibitem{DOV91} C. B. Dover,  Inst. Phys. Conf. Ser. \textbf{No 124-9} (1992) 421.
\bibitem{OEL91} W. Oelert,  Inst. Phys. Conf. Ser. \textbf{No 124-7} (1992) 307.
\bibitem{OEL93} W. Oelert, Nucl. Phys. \textbf{A 558} (1993) 73c. 
\bibitem{CES91} G. Cesari et al.,  Inst. Phys. Conf. Ser. \textbf{No 124-1} (1992) 13.
\bibitem{POL91} R.E. Pollock, Annu. Rev. Nucl. Part. Sci. \textbf{41} (1991) 357.
\bibitem{REI97} D. Reistad, Performance and Perspectives on CELSIUS, TSL-Note 97-31. 
\bibitem{MAI97} R. Maier, Nucl. Inst. \& Meth. \textbf{A 390} (1997) 1.
\bibitem{HAI92} J. Haidenbauer, K. Holinde, and J. Speth, Phys. Rev. \textbf{C 46} (1992) 2516.
\bibitem{BRA96} S. Brauksiepe et al., Nucl. Inst. \& Meth. \textbf{A 376} (1996) 397.
\bibitem{DOM97} H. Dombrowski et al., Nucl. Inst. \& Meth. \textbf{A 386} (1997) 228.
\bibitem{FIC62} W.J. Fickinger et al., Phys.Rev. \textbf{125} (1962) 2082.
\bibitem{FRA89} R. Frascaria, et. al., Nuovo Cimento \textbf{102A} (1989) 561.
\bibitem{RAN80} J. Randrup, and C.M. Ko, Nucl. Phys. \textbf{A 343} (1980) 519.
\bibitem{ZWE88} B. Sch\"urmann, and W. Zwermann, Mod. Phys. Lett. \textbf{A3} (1988) 251. 
\bibitem{FAL97} G. F\"aldt and C. Wilkin, Z. Phys. \textbf{A 357}(1997) 241.
\bibitem{SIB95} A. Sibirtsev, Phys. Lett. \textbf{B 359} (1995) 29; \\
                K. Tsushima, A. Sibirtsev and A.W. Thomas, Phys. Lett. \textbf{B 390} (1997) 29.
\bibitem{HAD97} HADRON'97, BNL, August 25 - 30 1997, \\ Conference Proceedings to be published.
\bibitem{CLO97} F. E. Close, hep-ph/9701290 - RAL-97-004 and see summary talk in ref.~\cite{HAD97}.
\bibitem{KLE97} E. Klempt, see summary talk in ref.~\cite{HAD97}.
\bibitem{JETSET} L. Bertolotto et al., Phys. Lett. \textbf{B 345} (1995) 325.
\end{thebibliography}
\end{document}